\documentstyle[11pt,epsf,psfig,axodraw]{article}

\def\Year{\expandafter\eatPrefix\the\year}
\newcount\hours \newcount\minutes
\def\monthname{\ifcase\month\or
January\or February\or March\or April\or May\or June\or July\or
August\or September\or October\or November\or December\fi}
\def\shortmonthname{\ifcase\month\or
Jan\or Feb\or Mar\or Apr\or May\or Jun\or Jul\or
Aug\or Sep\or Oct\or Nov\or Dec\fi}

\def\TimeStamp{\hours\the\time\divide\hours by60%
\minutes -\the\time\divide\minutes by60\multiply\minutes by60%
\advance\minutes by\the\time%
${\rm \shortmonthname}\cdot   \if\day<10{}0\fi\the\day\cdot   \the\year%
\qquad\the\hours:\if\minutes<10{}0\fi\the\minutes$}

\newskip\humongous \humongous=0pt plus 1000pt minus 100pt
\def\caja{\mathsurround=0pt}
\def\eqalign#1{\,\vcenter{\openup1\jot \caja
       \ialign{\strut \hfil$\displaystyle{##}$&$
        \displaystyle{{}##}$\hfil\crcr#1\crcr}}\,}
\newif\ifdtup

\newcounter{eqnumber}[section]
\renewcommand{\theeqnumber}{\thesection.\arabic{eqnumber}}
\def\equn{\refstepcounter{eqnumber}
\eqno({\rm \theeqnumber})
}

\newbox\charbox
\newbox\slabox
\def\s#1{{      
        \setbox\charbox=\hbox{$#1$}
        \setbox\slabox=\hbox{$/$}
        \dimen\charbox=\ht\slabox
        \advance\dimen\charbox by -\dp\slabox
        \advance\dimen\charbox by -\ht\charbox
        \advance\dimen\charbox by \dp\charbox
        \divide\dimen\charbox by 2
        \raise-\dimen\charbox\hbox to \wd\charbox{\hss/\hss}
        \llap{$#1$}
}}

\def\spa#1.#2{\left\langle#1\,#2\right\rangle}
\def\spb#1.#2{\left[#1\,#2\right]}
\def\lor#1.#2{\left(#1\,#2\right)}

\catcode`@=11  

\def\as#1{a_{\sigma(#1)}}
\def\sig#1{\sigma(#1)}

\def\Tr{\, {\rm Tr}}

\def\eps{\epsilon}

\def\pol{\eps}

\def\lsl{\not{\hbox{\kern-2.3pt $\ell$}}}
\def\ksl{\not{\hbox{\kern-2.3pt $k$}}}

\def\spa#1.#2{\left\langle#1\,#2\right\rangle}
\def\spb#1.#2{\left[#1\,#2\right]}
\def\lor#1.#2{\left(#1\,#2\right)}

\def\sand#1.#2.#3{%
  \left\langle\smash{#1}{\vphantom1}\right|{#2}%
  \left|\smash{#3}{\vphantom1}\right\rangle}
\def\sandp#1.#2.#3{%
  \left\langle\smash{#1}{\vphantom1}^{-}\right|{#2}%
  \left|\smash{#3}{\vphantom1}^{+}\right\rangle}
\def\sandpp#1.#2.#3{%
  \left\langle\smash{#1}{\vphantom1}^{+}\right|{#2}%
  \left|\smash{#3}{\vphantom1}^{+}\right\rangle}
\def\sandmm#1.#2.#3{%
  \left\langle\smash{#1}{\vphantom1}^{-}\right|{#2}%
  \left|\smash{#3}{\vphantom1}^{-}\right\rangle}
\def\sandpm#1.#2.#3{%
  \left\langle\smash{#1}{\vphantom1}^{+}\right|{#2}%
  \left|\smash{#3}{\vphantom1}^{-}\right\rangle}
\def\sandmp#1.#2.#3{%
  \left\langle\smash{#1}{\vphantom1}^{-}\right|{#2}%
  \left|\smash{#3}{\vphantom1}^{+}\right\rangle}

\def\Atree{A^{\rm tree}}

\def\dlips{dLIPS}

\def\beq{\begin{equation}}
\def\eeq{\end{equation}}
\def\beqa{\begin{eqnarray}}
\def\eeqa{\end{eqnarray}}

\begin{document}

\begin{titlepage}

\begin{flushright}

SWAT-03-332 \\

\end{flushright}

\vskip 2.cm

\begin{center}
\begin{Large}
{\bf Ultra-Violet Infinities and Counterterms 
in Higher Dimensional Yang-Mills}

\vskip 2.cm

\end{Large}

\vskip 2.cm

{\large 
David C. Dunbar 
and Nicolaus W. P. Turner $^{1}$}

\vskip 0.5cm

{\it Department of Physics, \\
University
    of Wales Swansea, 
\\ Swansea, SA2 8PP, UK }

\vskip 5.0cm

\begin{abstract}
In this letter we investigate the ultra-violet behaviour of 
four-point one-loop gluon amplitudes in dimensions greater than four
coupled to various particles types.
We discuss the structure of the counterterms and 
their inherent symmetries.

\end{abstract}

\vskip 1.0cm

\end{center}

\vfill
\noindent\hrule width 3.6in\hfil\break
%
%
${}^1$Research supported by PPARC \hfil\break

\end{titlepage}

\section{Introduction}

\noindent
In this letter we study the ultra-violet behaviour of higher
dimensional Yang-Mills theories. In four dimensions, 
Yang-Mills theory~\cite{YangMills} has
a dimensionless coupling constant 
and is in fact renormalisable~\cite{Renormalisation}. 
However, for $D>4$ the coupling constant has 
dimensions,
$$
[ g^2] = { (D-4) }
\equn
$$
and the presence of any 
ultra-violet infinities
would potentially render the theory
non-renormalisable.

\noindent
We shall examine the counterterms for pure Yang-Mills and Yang-Mills
coupled to a range of matter contents.  In general, infinities in
gluon scattering amplitudes are removed by counterterms of the form
$$
D^m F^n
\equn
$$
\noindent
(where the indices on the field strength $F_{a b}$ have been suppressed).
By determining the counterterms we aim
to gain insight into the structures needed to regulate theories with
dimensionfull coupling constants. This is in many ways the opposite
approach from starting with a very symmetric theory at high energies
and taking the low energy limit. Instead we hope to ``rediscover''
structures such as superstrings~\cite{GSW} by extrapolating to high
energies.  In some senses, the behaviour of Yang-Mills in $D > 4$
mirrors that of gravity in $D> 2$ and we hope to extend our
investigations to the case of gravity~\cite{DunbarTurnerB}.

We work within a dimensional reduction prescription~\cite{DimReg}, in
which the one-loop amplitudes are finite in odd dimensions.  Thus in
even dimensions, at one-loop, there is the possibility of divergences
in the amplitudes. For dimensions $D=6,8$ and $10$ we find the
counterterm structure and for even dimensions where $D >10$
we find an illustrative amplitude containing a non-vanishing
divergence.  Since for higher dimensions the infinities are, in
general, greater than logarithmic the structure of the counterterms is
very much dependent on the regularisation scheme and, in fact,
counterterms evaluated using a cut-off regularisation scheme would be
very different. Since dimensional regularisation manifestly
preserves the gauge symmetry of the amplitude, the counterterm structure
manifestly exhibits gauge invariant features.

We shall determine the counterterms by the calculation of on-shell
physical amplitudes. In a gauge theory the two- and three-point
functions vanish on-shell and hence will not determine the counterterm
structure and we must evaluate 
four-point amplitudes.
(Thus we are effectively only sensitive to
divergences up to $D^m F^n$ with $n\leq 4$.)

\section{Organisation of the Amplitudes}

Although we are interested in amplitudes in $D > 4$, we can still use
some of the powerful techniques used to evaluate amplitudes in
$D=4$~\cite{QCDReview}. 
In particular, we shall use a form of 
``spinor-helicity''~\cite{SpinorHelicity},
``color-ordering''~\cite{TreeColor,BKLoopColor}
 and a supersymmetric organisation of particle type.
Organising amplitudes carefully according to helicity, color and spin
can be termed ``total quantum number management''~\cite{LanceTasi}.

\noindent
{\bf Spinor Helicity: }
Spinor helicity is principally a four dimensional concept where the
polarisation vectors $\epsilon_{\mu}$ are realised as
combinations \cite{SpinorHelicity} of Weyl spinors $\vert k^{\pm} \rangle$,
$$
\pol^{+}_\mu (k;q) =
{\sandmm{q}.{\gamma_\mu}.k
\over  \sqrt2 \spa{q}.k} \hskip 2 cm
\pol^{-}_\mu (k;q) =
{\sandpp{q}.{\gamma_\mu}.k
\over \sqrt{2} \spb{k}.q}
\equn
$$
where $k$ is the gluon momentum and $q$ is an arbitrary null
`reference momentum' which drops out of the final gauge-invariant
amplitudes.  The plus and minus labels on the polarization vectors
refer to the gluon helicities and we use the notation
$\langle ij \rangle\equiv  \langle k_i^{-} \vert k_j^{+} \rangle\, ,
[ij] \equiv \langle k_i^{+} \vert k_j^{-} \rangle$.
These spinor products are anti-symmetric and satisfy
$\spa{i}.j \spb{j}.i = 2 k_i \cdot k_j \equiv s_{ij}$. 
For four-point amplitudes we will use the usual
Mandelstam variables $s= s_{12}$, $t= s_{14}$ and $u= s_{13}$.

We can use four dimensional helicity provided we identify a
suitable four dimensional subspace of the $D$ dimensions.  If we have a
four-point amplitude, momentum conservation implies we can use the three
independent 3-momenta plus time to define a four dimensional subspace.
In this frame the momenta of the scattered particles lie exclusively
in the four dimensional hyperspace.  Defining
$$
x^a =( x^{\mu} ; x^I )
\equn
$$
where there are $D-4$ coordinates $x^I$.
With this choice of coordinates
$$
k^I_i =0
\equn
$$
for the four external momenta, $k_i$, and
we can choose the helicity vectors $\eps_a$ to be of
two types: $\eps^{\pm}_a$ and $\eps^I_a$
$$
\eqalign{
\eps^{\pm}_a &= ( \eps^{\pm}_{\mu} ; 0 )
\cr
\eps^I_a &= ( \ 0 \ ; 0, \dots ,0,1,0, \dots ,0 )
\cr}
\equn
$$ 
which provide $D-2$ independent polarisation vectors.

\noindent{\bf Color Ordering:}
One-loop $SU(N_c)$ gauge theory amplitudes can be written in 
terms of independent color-ordered partial amplitudes multiplied 
by an associated color structure~\cite{TreeColor,BKLoopColor}.
The decomposition of the four-point one-loop gluon amplitude
(with adjoint particles in the loop) is
$$
\eqalign{
{\cal A}_4 (\{ a_i, k_i, \pol_i\}) =
 \sum_{\sigma} N_c \, & \Tr(T^{\as1}   T^{\as2} T^{\as3} T^{\as4} )
\ A_{4;1}(\sig1,\sig2,\sig3,\sig4)
\cr
&+  \sum_{\rho}
\Tr(T^{a_{\rho(1)}} T^{a_{\rho(2)}} )
\Tr(T^{a_{\rho(3)}} T^{a_{\rho(4)}} ) 
\ A_{4;3}(\rho (1),\rho (2);\rho (3),\rho (4))
\cr}
\equn
$$
where $a_i$, $k_i$ and $\epsilon_i$ are respectively the color index,
momentum and polarisation of the $i^{\rm th}$ external gluon. We have
also abbreviated the arguments of the `partial amplitudes', $A_{n;j}$,
by the labels $i$ of the legs and the $T^{a_i}$ are fundamental
representation matrices, normalized so that $\Tr(T^a T^b) =
\delta^{ab}$.  The sums over $\rho$ and $\sigma$ include all
non-cyclic permutations of the indicies $\sigma(i)$ and $\rho(i)$ which
leave the color trace structure invariant.  The structure for
any number of legs is similar, with no more than two color traces
appearing in each term (at one-loop).  String theory suggests, and it
has been proven in field theory, that the $A_{n;j>1}$ may be obtained
from $A_{n;1}$ by an appropriate permutation sum~\cite{BKLoopColor,Cutting2},
$$
 A_{n;j}(1,2,\ldots,j-1;j,j+1,\ldots,n)\ =\
 (-1)^{j-1} \sum_{\sigma\in COP\{\alpha\}\{\beta\}} A_{n;1}(\sigma)
\equn
$$
where $\alpha_i \in \{\alpha\} \equiv \{j-1,j-2,\ldots,2,1\}$,
$\beta_i \in \{\beta\} \equiv \{j,j+1,\ldots,n-1,n\}$,
and $COP\{\alpha\}\{\beta\}$ is the set of all
permutations of $\{1,2,\ldots,n\}$ with $n$ held fixed
that preserve the cyclic
ordering of the $\alpha_i$ within $\{\alpha\}$ and of the $\beta_i$
within $\{\beta\}$, while allowing for all possible relative orderings
of the $\alpha_i$ with respect to the $\beta_i$.

Thus, we need only consider the $A_{n;1}$, for they contain the
information necessary to reconstruct the full one-loop amplitude, and
any identity proven for the $A_{n;1}$ extends automatically to the
full amplitude. From here on we shall concentrate upon $A_{n;1}$
(often abbreviated to $A_{n}$).

Using spinor helicity, 
the color ordered amplitudes $A_4(1,2,3,4)$ can be organised according to the 
helicity of the external states which, by convention,  we take to be 
outgoing. 
For $D=4$, there are only four independent
amplitudes,
$$
A_4(1^+,2^+,3^+,4^+), \;\; 
A_4(1^-,2^+,3^+,4^+), \;\; 
A_4(1^-,2^-,3^+,4^+), \;\; 
A_4(1^-,2^+,3^-,4^+) \;\;
\equn
$$
the others being obtained by conjugation or cyclic permutation.
In $D>4$ we also have to consider
$$
A_4(1^+,2^+,3^I,4^I), \;\; 
A_4(1^+,2^I,3^+,4^I), \;\; 
A_4(1^-,2^+,3^I,4^I), \;\; 
A_4(1^-,2^I,3^+,4^I) \;\;
\equn
$$ 
and
$$
A_4(1^I,2^I,3^J,4^J), \;\; 
A_4(1^I,2^J,3^I,4^J), \;\; 
A_4(1^I,2^I,3^I,4^I) \;\;
\equn
$$ 
\noindent
where $I \neq J$.

\noindent{\bf Particle Content:}  Loop amplitudes 
depend upon the particle content of the theory and for gauge theories
 these depend upon both the spin and gauge representation of
the particles.  To determine the total ultra-violet structure we must determine
these different contributions. We concentrate upon particles
which lie in the adjoint of the gauge group and 
consider the contributions from three particle types: complex scalar,
fermion, and vector, which we denote by $A_{n;1}^{[J]}$ where $J$ may be $S$,
$F$ or $V$. We draw upon lessons from string-based techniques for computing
one-loop amplitudes in gauge theories~\cite{Long} which
reveal that the gluon amplitudes
are most simply calculated~\cite{Decomposition} by evaluating
the contributions from different supersymmetry
multiplets. In $D=4$, these contributions are those from a $N=4$ supermultiplet
and from a $N=1$ matter multiplet,
$$
\eqalign{
A_{n;1}^{N=4}\ &\equiv\
A_{n;1}^{[V]} + 4A_{n;1}^{[F]}+3 A_{n;1}^{[S]}
\cr
A_{n;1}^{N=1\ {\rm chiral}}\ &\equiv\
A_{n;1}^{[F]}\
+A_{n;1}^{[S]}
\cr}
\equn
$$
Experience shows that these are considerably easier to calculate than
$A_{n;1}^{[V]}$ and $A_{n;1}^{[F]}$ and from these supersymmetric
contributions and $A_{n;1}^{[S]}$ we can reconstruct,
$$
\eqalign{
A_{n;1}^{[V]} &= A_{n;1}^{N=4}-4 A_{n;1}^{N=1\,{\rm chiral}} + A_{n;1}^{[S]} \cr
A_{n;1}^{[F]} &= A_{n;1}^{N=1\ {\rm chiral}} - A_{n;1}^{[S]}
\cr}
\equn
$$
For $4\leq D\leq 10$ similar supersymmetric decompositions are very useful, but 
the exact form is dimension dependent due to the changing nature of the spinors
involved, however one combination is always the dimensional reduction of the
$D=10, N=1$ theory.
The contribution from this multiplet has been calculated
previously (using the low energy limit of string theory)~\cite{GSB} and the
final result in this case is particularly
simple; the amplitude is proportional to $st\Atree$ times a scalar
box integral. (The simplification in the maximally supersymmetric
case has allowed the calculation of the two-loop four-point amplitude
both for Yang-Mills~\cite{BRY} and Gravity \cite{BDDPR}.) The results 
for general particle type contain considerably more structure.

\def\denom{\spa1.2\spa2.3\spa3.4\spa4.1}

\section{Evaluating Amplitudes}

There are a variety of techniques for calculating loop amplitudes,
often more efficient than a Feynman diagram approach.  
In our calculations, we illustrate the use of two quite different
alternates to Feynman diagrams.

Firstly, we use Cutkosky 
cutting techniques~\cite{Cutting1,Cutting2,Cutting3}.  
The
optical theorem leads to the Cutkosky cutting rules in field theory
and it is possible to use these rules to determine amplitudes {\it
provided} one evaluates the cuts to ``all orders in
$\epsilon$''. (This is within the context of dimensional
regularisation where amplitudes are evaluated in $D=2 N -2\epsilon$.)
These all-$\epsilon$ results allow a complete reconstruction of the
amplitude and by analytically continuing in dimension one can obtain
the amplitudes for a range of dimensions.

Secondly, we use the Bern-Kosower rules for
evaluating QCD amplitudes~\cite{Long} which arose from the low energy
limit of string theory amplitudes.  In conventional field theory they
have been shown to be related to mixed gauge choices~\cite{Mapping} and
also to the ``World-line formalism''~\cite{WorldLine}. Since String
theory exists most naturally in $D=10$ or $D=26$, the rules are
trivially adapted to $D>4$.

\noindent 
{\bf From Cutting:} Amplitudes with purely four dimensional helicities
have been calculated previously \cite{BernMorgan,BernDelDucaSchmidt},
so for our first example, let us consider the amplitude 
$A^{\rm 1-loop}_4(1^I ,2^J ,3^I ,4^J)$. We  
 illustrate this technique by considering the cut in this amplitude in the
$s$-channel with all the gluons out-going.  According to the Cutkosky
rules it is given by
$$
 \left. -i \, {\rm Disc}~A^{\rm 1-loop}_4(1^I ,2^J ,3^I ,4^J)
\right|_{s-cut} \ =
\int\dlips\sum_{ {\rm internal}\atop {\rm states,s}}
\,\Atree_4(-\ell_1^s,1^I ,2^J ,\ell_2^s)\,\Atree_4(-\ell_2^s, 3^I ,4^J ,\ell_1^s )
\equn
\label{FourPtCut}
$$
as shown in figure~1. 

\begin{center}
\begin{picture}(300,120)(0,0)
\DashLine(150,15)(150,105){3}
\ArrowLine(120,30)(120,90)
\ArrowLine(120,90)(147,97)
\ArrowLine(153,97)(180,90)
\ArrowLine(180,90)(180,30)
\ArrowLine(180,30)(153,23)
\ArrowLine(147,23)(120,30)
\Gluon(120,90)(95,115){4}{3}
\Gluon(120,30)(95,5){4}{3}
\Gluon(180,90)(205,115){4}{3}
\Gluon(180,30)(205,5){4}{3}
\Text(95,5)[rt]{$1^I$}
\Text(95,115)[rb]{$2^J$}
\Text(205,115)[lb]{$3^I$}
\Text(205,5)[lt]{$4^J$}
\LongArrowArcn(150,80)(30,120,60)
\LongArrowArcn(150,40)(30,300,240)
\Text(150,115)[cb]{$\ell_2^s$}
\Text(150,5)[ct]{$\ell_1^s$}
\end{picture}
\vskip 0.5 cm
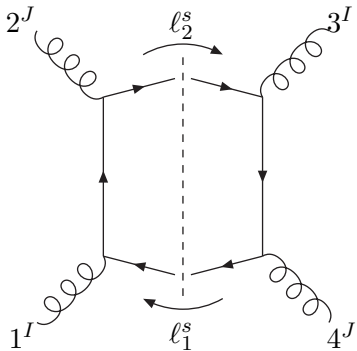
{\rm Figure 1: The s-channel cut}
\label{TreeRelationFigure}
\end{center}

To use the cutting rules to determine amplitudes precisely, we must be
careful in evaluating the tree amplitudes. Dimensional regularisation
applies to the internal loop momentum meaning that internal and
external legs are evaluated in different dimensions. For our case this
means the external legs lie in $2N$ dimensions and the internal legs
lie in $D=2N-2 \eps$ dimensions. (For calculational purposes we 
have let the $2N$ dimensional external momenta define a four dimensional
hyperspace.) 

In this example we will  consider a complex
scalar within the loop and
with these definitions, 
the tree amplitude for two external gluons and two internal 
complex scalars is
$$
\Atree_4 ( -\ell_1^{},1^I ,2^J ,\ell_2^{} )
= -i \, (\sqrt{2} g )^2 \, 
{l_1^I \, l_1^J  \over (l_1 -k_1)^2 }
\equn
$$
where $l_1^I$ denotes the $I^{\rm th}$  component of the loop
momentum $l_1$. Thus
$$
\eqalign{
-i \, {\rm Disc}~A^{\rm 1-loop}_4(1^I ,2^J ,3^I ,4^J) \Big|_{s-cut}
&=-2 \, (\sqrt{2} g )^4
 \int\dlips {(l_1^I)^2 \, (l_1^J)^2  \over (l_1 -k_1)^2 (l_1 +k_4)^2 }
\cr
&=2 \, (\sqrt{2} g )^4
\left.  \int { d^Dl_1 \over (2\pi)^D }
{(l_1^I)^2 \, (l_1^J)^2  \over l_1^2 (l_1 -k_1)^2 l_2^2 (l_1 +k_4)^2 } \right|_{s-cut}
\cr}
\equn
$$
This s-channel expression is also correct for the t-channel cut and so we can 
determine the full amplitude to be
$$
A_4^{[S]}(1^I, 2^J, 3^I, 4^J) =
 2 i \, {(\sqrt{2} g)^4 \over (4 \pi)^{D/2}} \ I_4^{D}[(l_1^I)^2 \, (l_1^J)^2]
\equn
\label{massless_all-plus}
$$
The $(l_1^I)^2$ effectively shifts the dimension of the integral and so
for general dimension, $D$, we have,
$$
A_4^{[S]}(1^I, 2^J, 3^I, 4^J) = i \, {(\sqrt{2} g)^4 \over (4 \pi)^{D/2}}
 \ { 1 \over 2 } \ I_4^{D+4}
\equn
$$
\noindent
where the $n$-point scalar one-loop integral in $D$ dimensions is
$$
{ I}_{n}^{D} = i \ (-1)^{n+1} \ (4 \pi)^{D/2}
\int \frac{d^D p}{(2 \pi)^D} 
\frac{1}{p^2 (p-k_{1})^2 (p-k_{1}-k_{2})^2 \dots (p-k_{1}-k_{2}-\dots-k_{n-1})^2}
\equn
$$
This amplitude is quite similar to the amplitude 
$A_4^{[S]}(1^+, 2^+, 3^+, 4^+)$~\cite{DimShift}
with
$$
A_4^{[S]}(1^+, 2^+, 3^+, 4^+)
=- { st \over \denom }{ (D-4)(D-2) \over 4}
A_4^{[S]}(1^I, 2^J, 3^I, 4^J)
\equn
$$
The amplitude  $A_4^{[S]}(1^I, 2^J, 3^I, 4^J)$
will contain divergences in even dimensions and
extracting these for this amplitude gives
$$
\eqalign{
D=6: & -i \, { (\sqrt{2} g)^4 \over (4\pi)^3 \ \eps }
 \ {u \over 240 }
\cr
D=8: & +i \, {  (\sqrt{2} g)^4\over (4\pi)^4 \ \eps }
 \ {( 2t^2+st+2s^2 ) \over 10080 }
\cr
D=10: & +i \, { (\sqrt{2} g)^4\over (4\pi)^5 \ \eps }
 \ {
 (3s^3+st^2+s^2t+3t^3) \over 362880 }
\cr
D=12: & +i \, { (\sqrt{2} g)^4\over (4\pi)^6 \ \eps }
\
{ (12 s^4 +3s^3t +2s^2t^2 +3st^3 +12 t^4) \over
39916800}
\cr
D=14: & +i \, { (\sqrt{2} g)^4\over (4\pi)^7 \ \eps }
\ {
(10s^5 +2s^4t +s^3t^2 +s^2t^3+ 2st^4 +10t^5) \over 1037836800}
\cr}
\equn
$$
These clearly show that this amplitude has a non-vanishing divergence
for even dimensions where $6 \leq D \leq 14$. 
Although, we do not explicitly present them,
the amplitude has a non-vanishing divergence
for all even dimensions, $D \geq 16$.

\noindent
{\bf From String Based Rules:} The string based rules
provide a compact mechanism for obtaining the Feynman parameter
polynomials in one-loop integrals.
For example, examining the amplitude
$A_4(1^-, 2^+, 3^+, 4^+)$ using these rules, 
we find the contribution to the one-loop amplitude in $D=4$
due to a complex scalar is~\cite{Long,QCDReview}
$$
A_4^{[S]}(1^-, 2^+, 3^+, 4^+)
=  -i \, (\sqrt{2} g) ^4 \
{ \spb2.4^2 \over \spb1.2 \spa2.3 \spa3.4\spb4.1 } \ {s^2 t \over 2} \
\left( {\cal D}_4(s,t)^{(a)} +{\cal D}_3(s)^{(b)}
\right)
\equn
$$
where the Feynman parameter integrals are
$$
\eqalign{
{\cal D}_4(s,t)^{(a)} &=
{\Gamma(4-D/2)\over (4\pi)^{D/2}}
\int_0^1 \prod_{i=1}^4 da_i \ \delta\left(1-\sum_{i=1}^4
a_i\right)
{- (a_1+a_2)a_3^2a_4 \over(-sa_1a_3-ta_2a_4)^{4-D/2}}
\cr
{\cal D}_3(s)^{(b)} &= 
{\Gamma(3-D/2)\over (4\pi)^{D/2}} 
{ 1 \over s } \int_0^1  da_1da_3da_4 \ \delta\left(1-\sum_i
a_i\right)
{ a_1a_3a_4
\over(-s a_1a_3)^{3-D/2}}
\cr}
\equn
$$
The String based rules for $D=4$ arise from a reduction of $D=10$ 
string theory, so the rules can be trivially adapted to any $D\leq 10$ and
the Worldline formalism suggests they would be valid in any dimension. 
The Feynman parameter polynomial will be identical in all dimensions and so
evaluating the loop integrals we have the following
overall divergences

$$
\eqalign{
D=6:  & -i \, { (\sqrt{2} g) ^4 \over (4\pi)^3 \ \eps }
\ { \spb2.4^2 \over \spb1.2 \spa2.3 \spa3.4\spb4.1 }
\ { s t \over 240 }
\cr
D=8:  & \ \ \ \ \ \ 0
\cr
D=10:  & +i \, { (\sqrt{2} g) ^4 \over (4\pi)^5 \ \eps }
\ { \spb2.4^2 \over \spb1.2 \spa2.3 \spa3.4\spb4.1 }
\ { s^2 t^2 \over 60480 }
\cr}
\equn
$$

For the choices of helicity, $A_4(1^+,2^+,3^+,4^+)$ and
$A_4(1^-,2^+,3^+,4^+)$, the supersymmetric contributions
vanish in any dimension $D \leq
10$. In $D=4$ this is shown using supersymmetric Ward
identities~\cite{SWI} and for $D > 4$ the supersymmetric algebra contains
the $D=4$ superalgebra as a subalgebra and hence these contributions
will also vanish.

\vskip 0.3 truecm 

\noindent {\bf Counterterms in ${\bf D=6} $:} Counterterms for
one-loop calculations may be of the form $F^3$ or $D^2F^2$.  By using
the equations of motion and the Bianchi identities the terms quadratic
in $F$ may be eliminated \cite{Morozov} leaving a single possible
counterterm
$$
\Tr \left(
F_a^{\ b} F_b^{\ c} F_c^{\ a} \right)
\equn
$$
The coefficient of
this single term could be determined from any of the (non-zero)
one-loop amplitudes and could depend upon the particle content of the
theory in a non-trivial way. However, this simplifies as can be seen
by looking at the amplitude $A_4(1^+, 2^+, 3^+, 4^+)$. As discussed
above, this amplitude vanishes in any supersymmetric theory, and since
the $F^3$ counterterm is non-vanishing for this amplitude its
coefficient must vanish in any supersymmetric theory.  (The amplitudes
corresponding to this counterterm in $D=6$ appear when considering
higher derivative theories in $D=4$~\cite{DixonShadmi}.)  This
simplifies the amplitude considerable when using a supersymmetric
decomposition where the contributions from the supersymmetric
multiplets vanish.

In $D=6$ we can consider the $N=2$ multiplet (which is the
reduction of the $D=10,N=1$ theory) which contains one vector, two Weyl fermions 
and two complex scalars, so that,
$$
A^{N=2}_4
=A^{[V]}_4
+2A^{[F]}_4
+2A^{[S]}_4
\equn
$$
There is also the $N=1$ vector multiplet with  one vector and one 
Weyl fermion,
$$
A^{N=1}_4
=A^{[V]}_4
+A^{[F]}_4
\equn
$$
Inverting these relationships gives,
$$
\eqalign{
A^{[V]}_4
&=-A^{N=2}_4
+2A^{N=1}_4
+2A^{[S]}_4
\cr
A^{[F]}_4
&=A^{N=2}_4
-A^{N=1}_4
-2A^{[S]}_4
\cr}
\equn
$$
Although the amplitudes $A^{N=2}_4$ and $A^{N=1}_4$ are non-vanishing
for general helicities they have no ultra-violet infinities by the
above argument. If our theory has $N_v$ vector particles, $N_f$ fermions and
$N_s$ complex scalars, the counterterm will be proportional to
$$
2N_v -2N_f +N_s
\equn
$$
or equivalently 
$$
N_B-N_F
\equn
$$
where $N_B$ is the total number of bosonic degrees of freedom and $N_F$
is the total number of fermionic degrees of freedom.  Therefore, in $D=6$, it is
possible to determine the entire counterterm structure by examining the contributions
to the amplitudes from scalar particles circulating in the loop.

Using the string based rules, or other techniques, one can
obtain the complete set of amplitudes and extract the infinities, giving
$$
\eqalign{
A^{[S]}_4(1^+, 2^+, 3^+, 4^+) \Big|_{1 \over \eps}
& =  +i \, { (\sqrt{2} g )^4 \over (4\pi)^3 \ \eps }
\ { 1 \over \spa1.2\spa2.3\spa3.4\spa4.1 }
\ {s t u \over 120} \nonumber \\
\cr
A^{[S]}_4(1^-, 2^+, 3^+, 4^+)\Big|_{1 \over \eps}
& =  -i \, { (\sqrt{2} g )^4 \over (4\pi)^3 \ \eps }
\ { \spb2.4^2 \over \spb1.2 \spa2.3 \spa3.4\spb4.1 }
\ { s t \over 240} 
\cr
A^{[S]}_4(1^-, 2^-, 3^+, 4^+)\Big|_{1 \over \eps}
& =  \ \ \ \ 0 
\cr
A^{[S]}_4(1^-, 2^+, 3^-, 4^+)\Big|_{1 \over \eps}
& =  \ \ \ \ 0 
\cr
A^{[S]}_4(1^+, 2^+, 3^I, 4^I)\Big|_{1 \over \eps}
& =  -i \, { (\sqrt{2} g )^4 \over (4\pi)^3 \ \eps }
\ { \spa3.4^2 \over \spa1.2\spa2.3\spa3.4\spa4.1 } \ { t u \over 240} 
\cr
A^{[S]}_4(1^+, 2^I, 3^+, 4^I)\Big|_{1 \over \eps}
& =  -i \, { (\sqrt{2} g )^4 \over (4\pi)^3 \ \eps }
\ { \spa2.4^2 \over \spa1.2\spa2.3\spa3.4\spa4.1 } \ { s t \over 240 } 
\cr
A^{[S]}_4(1^-, 2^+, 3^I, 4^I)\Big|_{1 \over \eps}
& =  \ \ \ \ 0 
\cr
A^{[S]}_4(1^-, 2^I, 3^+, 4^I)\Big|_{1 \over \eps}
& =  \ \ \ \ 0 
\cr
A^{[S]}_4(1^I, 2^I, 3^J, 4^J)\Big|_{1 \over \eps}
& =  -i \, { (\sqrt{2} g )^4 \over (4\pi)^3 \ \eps }
 \ { u  \over 240 } 
\cr
A^{[S]}_4(1^I, 2^J, 3^I, 4^J)\Big|_{1 \over \eps}
& =  -i \, { (\sqrt{2} g )^4 \over (4\pi)^3 \ \eps }
\ { u  \over 240 }
\cr
A^{[S]}_4(1^I, 2^I, 3^I, 4^I) \Big|_{1 \over \eps}
& =  -i \, { (\sqrt{2} g )^4 \over (4\pi)^3 \ \eps }
\ { u  \over 80  } 
\cr}
\equn
$$
Comparing the one-loop infinity from any of the non-zero amplitudes
with an amplitude calculated from  the $F^3$ term we
can determine the $D=6$ counterterm structure to be
$$
i \, (N_B-N_F) \ { (\sqrt{2} g )^3 \over (4\pi)^3 \epsilon} \
{ N_c \over 720 }
\Tr \left(
F_a^{\ b} F_b^{\ c} F_c^{\ a} \right)
\equn
$$
This is for adjoint particles only, if we have $n_f$ representations
of fundamental particles the factor $N_c$ is replaced by $n_f$.
The counterterm obviously vanishes in a supersymmetric theory or in
any theory where the number of bosonic and fermionic degrees of
freedom are equal and then the theory is one-loop finite.

\vskip 0.3 truecm

\noindent {\bf Counterterms in ${\bf D=8} $:}
The $D=8$ case is more complex than that of $D=6$.
A general counterterm will be a dimension eight operator. 
These can be of the symbolic forms
$$
D^4 F^2 \hskip 1cm D^2 F^3 \hskip 1cm F^4
\equn
$$
The form of the counterterm can be changed by using the equations of motion. 
In general the equations of motion will be of the form
$$
D^{a} F_{ab} = { \delta {\cal L}_{matter} \over  \delta A^{b} }
\equn
$$
where ${\cal L}_{matter}$ is the Lagrangian density for the matter
which is coupled to the gauge field and generally ${\delta {\cal
L}_{matter} / \delta A^{b} }$ is quadratic in the matter fields.
Counterterms differing by expressions involving $D^{a} F_{ab}$
will thus give the same purely gluonic amplitudes but will, in
general, give differing contributions to amplitudes involving at least
two matter fields. This means there are ambiguities in the purely
gluonic counterterms which cannot be resolved without the calculation
of amplitudes involving matter fields. Wherever possible we shall
use identities to replace terms with less than four field strengths.

For example, consider the $D^2F^3$ terms. These have been enumerated
in ref~\cite{ChoSimmons} for $SU(3)$ 
where they find three such operators, two of which involve $D^{a} F_{ab}$,
and so can be replaced using the equations of motion. 
The third of these does not involve the equations of motion directly
$$
O_3^{(8)} \sim \Tr \left( F_a^{\ b} F_b^{\ c} D^2 F_c^{\ a} \right) 
\equn
$$
However, using
$$
D^2 F_a^{\ b} = i \, \sqrt{2} g \, [F_a^{\ c} , F_c^{\ b} ] 
+ D^b D_c F_{a}^{\ c} - D_a D_c F^{bc} 
\equn
$$
this operator can be replaced by the combination
$$
i \, \sqrt{2} g \, \Tr \left( F_a^{\ b} F_b^{\ c} [F_c^{\ d}, F_d^{\ a}] \right)
= i \, \sqrt{2} g \, \Tr \left( F_a^{\ b}F_b^{\ c}F_c^{\ d}F_d^{\ a} \right)
- i \, \sqrt{2} g \,  \Tr \left( F_a^{\ b}F_b^{\ c}F_d^{\ a}F_c^{\ d} \right)
\equn
$$
Thus when considering physical four gluon amplitudes in $D=8$ it is
consistent to take as a basis of counterterms the four field strengths
operators.  In general, if possible, it makes good calculational sense
to choose counterterms which mirror the structure of the amplitudes.

There are eight possible counterterms of the form $F^4$,
$$
\eqalign{
G_1 &= \Tr \left( F_a^{\ b}F_b^{\ c}F_c^{\ d}F_d^{\ a} \right)
\cr
G_2 &=\Tr \left( F_a^{\ b}F_c^{\ d}F_b^{\ c}F_d^{\ a} \right)
\cr
G_3 &=\Tr \left( F_{ab}F^{ab}F_{cd}F^{cd} \right)
\cr
G_4 &= \Tr \left( F_{ab}F_{cd}F^{ab}F^{cd} \right)
\cr
G_5 &= \Tr \left( F_a^{\ b}F_b^{\ c} \right) \, \Tr \left( F_c^{\ d}F_d^{\ a} \right)
\cr
G_6 &=\Tr \left( F_a^{\ b}F_c^{\ d} \right) \, \Tr \left( F_b^{\ c}F_d^{\ a} \right)
\cr
G_7 &=\Tr \left( F_{ab}F^{ab} \right) \, \Tr \left( F_{cd}F^{cd} \right)
\cr
G_8 &= \Tr \left( F_{ab}F_{cd} \right) \ \ \Tr \, ( F^{ab}F^{cd} )
\cr}
\equn
$$
Of these four involve a double trace. These double trace counterterms
will not contribute to the leading in color $A_{n;1}$ but only to the
$A_{n;j>1}$. Since the $A_{n;j>1}$ are deducible from the $A_{n;1}$
we need not consider these double traces here. 
\footnote{The eight counterterms have been enumerated previously by
Morozov \cite{Morozov} using a non-color ordered formulation. His
tensors are linear combinations of ours. Specifically, for $SU(N_c)$;
$M_1 \equiv ( {\cal G}^2 )^2 = G_7$ ,
$M_2 \equiv ( {\cal G}\tilde {\cal G} )^2=4G_6-2G_8$, 
$M_3 \equiv ( {\cal G}^a {\cal G}^b)^2 =G_8 $, 
$M_4 \equiv ( {\cal G}^a \tilde {\cal G}^b)^2=4G_5-G_7-G_8$, 
$M_5 \equiv (d {\cal G} {\cal G} )^2 =-4 G_3 $,
$M_6 \equiv (d {\cal G} \tilde {\cal G} )^2=-16G_2+4(G_3+G_4)+\frac{8}{N_c}(2G_6-G_8)$, 
$M_7 \equiv ( d^{abm} {\cal G}_{pq}^a {\cal G}^b_{rs} ) ( d^{cdm} {\cal G}_{pq}^c {\cal G}^d_{rs} )
=-2(G_3+G_4)+\frac{4}{N_c}G_8$, 
$M_8 \equiv( d^{abm} {\cal G}_{pq}^a {\cal G}^b_{rs} ) 
( d^{cdm}\tilde {\cal G}_{pq}^c \tilde{\cal G}^d_{rs})=-8(G_1+G_2)+6G_3+2G_4+\frac{4}{N_c}(4G_5-G_7-G_8)$. 
In ref~\cite{ChoSimmons} there are only
six $F^4$ tensors since for  $SU(3)$ only six of these eight
are independent. }

Thus we consider a general 
counterterm to be of the form,
$$
c_1 G_1 +c_2 G_2 +c_3 G_3 +c_4 G_4
\equn
$$
Calculating with this general counterterm will give expressions for the
amplitudes. For example, in the all-plus case we have,
$$
\eqalign{
A_4(1^+, 2^+, 3^+, 4^+) &= -i \, { ( \sqrt{2} g)^4 \over (4 \pi)^4} 
\ { st \over \spa1.2\spa2.3\spa3.4\spa4.1 }
\ \left[ (c_2 + 2 c_3 + 4 c_4) (s^2  + t^2 ) - 2 (c_1 - c_2 - 4 c_4) s t
\right]
\cr}
\equn
$$
From this we can see that it will be necessary to evaluate more than a
single amplitude in order to determine the entire counterterm
structure.

For the different contributions circulating in the loop we can
consider the simple supersymmetric $D=8,N=1$ multiplet which consists
of a vector, a Weyl fermion and a single complex scalar,
$$
A^{N=1}_4
=A^{[V]}_4
+A^{[F]}_4
+A^{[S]}_4
\equn
$$
This is the only supersymmetric contribution, however, it is still
useful to separate a ``scalar'' part from the vector and fermion
contributions,
$$
\eqalign{
A^{[V]}_4
=3A^{[S]}_4
+A^{[V-S]}_4
\cr
A^{[F]}_4
=-4A^{[S]}_4
+A^{[F-S]}_4
\cr}
\equn
$$
We will calculate all four contributions; $A_4^{[S]}$, $A^{[V-S]}_4$,
$A^{[F-S]}_4$ and
$A^{N=1}_4$, although only three are independent since
$$
A^{N=1}_4= A^{[V-S]}_4+A^{[F-S]}_4
\equn
$$
Calculating the entire set of amplitudes, we
have infinities which are summarised
in table~1. We find that the different contributions to the counterterms are
$$
\eqalign{
G^{[S]} =& {( \sqrt{2} g)^4 \over (4 \pi)^4 \epsilon} \ { N_c \over 80640 }
\biggl( 4 G_1 + 24 G_2 + 34 G_3 + G_4 \biggr)
\cr
G^{[F]} =& {( \sqrt{2} g)^4 \over (4 \pi)^4 \epsilon} \ { N_c \over 5040 }
\biggl( 20 G_1 + 78 G_2 - 19 G_3 - 16 G_4 \biggr)
\cr
G^{[V]} =& {( \sqrt{2} g)^4 \over (4 \pi)^4 \epsilon} \ { N_c \over 26880 }
\biggl( 452 G_1 + 696 G_2 - 190 G_3 - 55 G_4 \biggr)
\cr
G^{N=1} =& {( \sqrt{2} g)^4 \over (4 \pi)^4 \epsilon} \ { N_c \over 192}
\biggl( 4 G_1 + 8 G_2 - 2 G_3 - G_4 \biggr)
\cr}
\equn
$$

%
\begin{table}[htb]
\hbox{
\def\tend{\cr \noalign{ \hrule}}
\def\t#1{\tilde{#1}}
\def\tw{\theta_W}
\vbox{\offinterlineskip
{
\hrule
\halign{
       &  \vrule#
        &\strut\hfil #\hfil\vrule
        &\strut\hfil #\hfil\vrule
        &\strut\hfil #\hfil\vrule
        &\strut\hfil #\hfil\vrule
        &\strut\hfil #\hfil\vrule
        &\strut\hfil #\hfil\vrule
       \cr
height15pt  &{\bf Amplitude}  & {\bf Overall Factor}   &
$\bf A_4^{[S]}$  & $\bf A_4^{N=1}$  & $\bf A_4^{[V-S]}$ & $\bf A_4^{[F-S]}$ &\tend
height20pt  & $A_4 (1^+, 2^+, 3^+, 4^+)$ &
$-i \, { ( \sqrt{2} g)^4 \over (4 \pi)^4 \ \eps} \ { 1 \over \spa1.2\spa2.3\spa3.4\spa4.1 } \ {s^2 t u \over 4} $     &
${ 2 s^2  + s t + 2 t^2 \over 420 su } $  &
$ 0 $ & 
$ 0$  &
$ 0$ &  \tend
height20pt  & $A_4 (1^-, 2^+, 3^+, 4^+)$  &
$-i \, { ( \sqrt{2} g)^4 \over (4 \pi)^4 \ \eps} \ { \spb2.4^2 \over \spb1.2 \spa2.3 \spa3.4\spb4.1 } \ {s^2 t \over 4} $   &
$0 $  &
$0$   &
$0$ &
$0$ &  \tend
height20pt  & $A_4 (1^-, 2^-, 3^+, 4^+)$  &
$-i \, { ( \sqrt{2} g)^4 \over (4 \pi)^4 \ \eps} \ { \spa1.2^4 \over \spa1.2 \spa2.3 \spa3.4\spa4.1 } \ {s t \over 4} $   &
$ { 1 \over 210} $  &
$ { 1 \over 6} $  &
$ { 1 \over 10} $  &
$ { 1 \over 15} $ & \tend
height 20pt  & $A_4 (1^-, 2^+, 3^-, 4^+)$
&  $ -i \, { ( \sqrt{2} g)^4 \over (4 \pi)^4 \ \eps} \ { \spa1.3^4 \over \spa1.2 \spa2.3 \spa3.4\spa4.1 } \ {s t \over 4} $  &
$ { 1 \over 630} $  &
$ { 1 \over 6} $  &
$ { 2 \over 15} $  &
$ { 1 \over 30} $  &  \tend
height20pt  & $A_4 (1^+, 2^+, 3^I, 4^I)$
& $-i \, { ( \sqrt{2} g)^4 \over (4 \pi)^4 \ \eps} \ { \spa3.4^4 \over \spa1.2 \spa2.3 \spa3.4\spa4.1 } \ {s t \over 2} $   &
$ - { 10 s + t \over 5040} $  &
$ 0 $  &
$ - { u \over 120} $  &
$ { u \over 120} $  & \tend
height20pt  & $A_4 (1^+, 2^I, 3^+, 4^I)$  &
$ \ i \, { ( \sqrt{2} g)^4 \over (4 \pi)^4 \ \eps} \ { \spa2.4^4 \over \spa1.2 \spa2.3 \spa3.4\spa4.1 } \ {s t \over 2} $   &
$ { u \over 2520} $  &
$ 0 $   &
$ { u \over 120} $  &
$ - { u \over 120} $ & \tend
height20pt  & $A_4 (1^-, 2^+, 3^I, 4^I)$  &
$-i \, { ( \sqrt{2} g)^4 \over (4 \pi)^4 \ \eps} \ { \spb1.2^2 \spa1.3^2 \over \spb1.2 \spa2.3 \spa3.4\spb4.1 } \ {t^2 \over 2 s} $   &
$ - { s \over 2520} $  &
$ - { s \over 12} $   &
$ - { 7 s \over 120} $  &
$ - { s \over 40} $ &\tend
height20pt  & $A_4 (1^-, 2^I, 3^+, 4^I)$  &
$-i \, { ( \sqrt{2} g)^4 \over (4 \pi)^4 \ \eps} \ { 1 \over \spb1.2 \spa2.3 \spa3.4\spb4.1 } \ { s^2 t^2  \over 2 u} $   &
$ - { u \over 1680} $  &
$ - { u \over 12} $  &
$ - { u \over 20} $  & 
$ - { u \over 30} $  & \tend
height20pt  & $A_4 (1^I, 2^I, 3^J, 4^J)$ &
$ i \, { ( \sqrt{2} g)^4 \over (4 \pi)^4 \ \eps} $   &
$ { 10 s^2 + 3 s t + 2 t^2 \over 10080}  $ &
$ - { t u \over 24} $ &
$ - { s^2 - 6 s t - 7 t^2 \over 240} $ &
$ { s^2 + 4 s t + 3 t^2 \over 240} $ & \tend
height20pt  & $A_4 (1^I, 2^J, 3^I, 4^J)$ &
$ i \, { ( \sqrt{2} g)^4 \over (4 \pi)^4 \ \eps} $   &
$ { 2 s^2 + s t + 2 t^2 \over 10080} $ &
$ - { s t \over 24} $ & 
$ { s^2 - 4 s t + t^2 \over 240} $    &
$ - { s^2 + 6 s t + t^2 \over 240} $ & \tend
height20pt  & $A_4 (1^I, 2^I, 3^I, 4^I)$ &
$ i \, { ( \sqrt{2} g)^4 \over (4 \pi)^4 \ \eps} $  &
$ { 2 s^2 + s t + 2 t^2 \over 1440} $  &
$ { s^2 + s t + t^2 \over 24} $  &
$ { 7 s^2 + 8 s t + 7 t^2 \over 240} $  &
$ { 3 s^2 + 2 s t + 3 t^2 \over 240} $  & \tend
}
}
}
}
\nobreak
\caption[]{The $1/ \epsilon$ infinities for the four-point one-loop gluon amplitudes in $D=8$.
\label{AmplitudesD8}
\smallskip}
\end{table}

\noindent
The $N=1$ contribution has been  calculated previously~\cite{GSB} and can be 
rewritten  as
$$
G^{N=1} = {( \sqrt{2} g)^4 \over (4 \pi)^4 \epsilon} \ { N_c \over 386} \ 
t_8 \cdot F^4
\equn
$$
where the tensor $t_8$ appears in several contexts related to string theory
and is defined in equation~9.A.18 of reference~\cite{GSW}.
These counterterms do not vanish for a supersymmetric theory, nor is there any
choice of $N_s$, $N_f$ and $N_v$ for which
$$
\left. \Bigl( N_s A^{[S]}_4 + N_v A^{[V]}_4 + N_f A^{[F]}_4 \Bigr) \right|_{1 \over \epsilon} = 0
\equn
$$

\vskip 0.3 truecm

\noindent {\bf Counterterms in ${\bf D=10} $:}
In general there are rather a lot of possible dimension ten operators 
which can act as counterterms. These can be of the form,
$$
D^6 F^2 
\hskip 1cm 
D^4 F^3
\hskip 1cm 
D^2 F^4
\hskip 1cm 
F^5
\equn
$$
Of these the $F^5$ terms will only contribute to amplitudes with at
least five gluons, so we will not determine them here.  As in the $D=8$
case there are ambiguities between the $D^4 F^3$ and $D^2 F^4$ terms
which cannot be resolved using only four gluon amplitudes.
Fortunately, as in $D=8$ it is possible to choose the counterterms to
be entirely of the form $D^2F^4$.

There are seven possible
single-trace linearly independent counterterms of the form $D^2 F^4$,

$$
\eqalign{
H_1 & = \Tr \left( D_{e} F_a^{\ b} D^{e} F_b^{\ c} F_c^{\ d} F_d^{\ a} \right)
\cr
H_2 & = \Tr \left( D_{e} F_a^{\ b} D^{e} F_c^{\ d} F_b^{\ c} F_d^{\ a} \right)
\cr
H_3 & = \Tr \left( D_{e} F_a^{\ b} D^{e} F_d^{\ a} F_b^{\ c} F_c^{\ d} \right)
\cr
H_4 & = \Tr \left( D_{e} F_{ab} D^{e} F^{ab} F_{cd} F^{cd} \right)
\cr
H_5 & = \Tr \left( D_{e} F_{ab} D^{e} F_{cd} F^{ab} F^{cd} \right)
\cr
H_6 & = \Tr \left( D_{e} F_{ab} D^{e} F_{cd} F^{cd} F^{ab} \right)
\cr
H_7 & = \Tr \left( D_{e} F_{a}^{\ b} D^{a} F_{b}^{\ c} F_{c}^{\ d} F_{d}^{\ e} \right)
\cr}
\equn
$$

For the different contributions circulating in the loop we can consider the simple
supersymmetric $D=10,N=1$ multiplet which consists of a vector and a Majorana-Weyl fermion,
$$
A^{N=1}_4
=A^{[V]}_4
+A^{[F]}_4
\equn
$$
This is the only supersymmetric contribution and again we extract
a ``scalar'' contribution,
$$
\eqalign{
A^{[V]}_4
=4A^{[S]}_4
+A^{[V-S]}_4
\cr
A^{[F]}_4
=-4A^{[S]}_4
+A^{[F-S]}_4
\cr}
\equn
$$
and calculate all four contributions; $A_4^{[S]}$, $A^{[V-S]}_4$,
$A^{[F-S]}_4$ and
$A^{N=1}_4$. The infinities for the entire set of amplitudes are
shown in tables~2 and 3 and these give rise to the following counterterms,

%
\begin{table}[htp]
\hbox{
\def\tend{\cr \noalign{ \hrule}}
\def\t#1{\tilde{#1}}
\def\tw{\theta_W}
\vbox{\offinterlineskip
{
\hrule
\halign{
       &  \vrule#
        &\strut\hfil #\hfil\vrule
        &\strut\hfil #\hfil\vrule
        &\strut\hfil #\hfil\vrule
        &\strut\hfil #\hfil\vrule
        &\strut\hfil #\hfil\vrule
        &\strut\hfil #\hfil\vrule
       \cr
height15pt  &{\bf Amplitude}  & {\bf Overall Factor}   &
$\bf A_4^{[S]}$  & $\bf A_4^{N=1}$  & $\bf A_4^{[V-S]}$ & $\bf A_4^{[F-S]}$ &\tend
height20pt  & $A_4 (1^+, 2^+, 3^+, 4^+)$ &
$ -i \, { ( \sqrt{2} g)^4 \over (4 \pi)^5 \ \eps} \ { 1 \over \spa1.2\spa2.3\spa3.4\spa4.1 } \ {s^2 t u \over 4} $     &
${ 3 s^3  + s^2 t + s t^2 + 3 t^3 \over 7560 su } $  &
$ 0 $ &
$ 0$  &
$0 $ &  \tend
height20pt  & $A_4 (1^-, 2^+, 3^+, 4^+)$  &
$ -i \, { ( \sqrt{2} g)^4 \over (4 \pi)^5 \ \eps} \ { \spb2.4^2 \over \spb1.2 \spa2.3 \spa3.4\spb4.1 } \ {s^2 t \over 4} $   &
$ - { t \over 15120} $  &
$0$   &
$0 $ &
$0 $ &  \tend
height20pt  & $A_4 (1^-, 2^-, 3^+, 4^+)$  &
$-i \, { ( \sqrt{2} g)^4 \over (4 \pi)^5 \ \eps} \ { \spa1.2^4 \over \spa1.2 \spa2.3 \spa3.4\spa4.1 } \ {s t \over 4} $   &
$ { 3 s + t \over 7560} $  &
$ -{ u \over 120} $  &
${  3 s + 5 t\over 840}$  &
$ {2 s +t\over 420 }$ & \tend
height 20pt  & $A_4 (1^-, 2^+, 3^-, 4^+)$
&  $ -i \, { ( \sqrt{2} g)^4 \over (4 \pi)^5 \ \eps} \ { \spa1.3^4 \over \spa1.2 \spa2.3 \spa3.4\spa4.1 } \ {s t \over 4} $  &
$ -{ u \over 15120} $  &
$ -{ u \over 120} $  &
$-{17u \over 2520} $  &
$ -{u \over 630}$  &  \tend
height20pt  & $A_4 (1^+, 2^+, 3^I, 4^I)$
& $-i \, {( \sqrt{2} g)^4 \over (4 \pi)^5 \ \eps} \ { \spa3.4^4 \over \spa1.2 \spa2.3 \spa3.4\spa4.1 } \ {s t \over 2} $   &
$ - { 33 s^2 + 5 s t - 4 t^2 \over 181440} $  &
$ 0 $  &
$ {2 s^2  + s t + 2 t^2 \over 2520}$  &
$ -{2 s^2  + s t + 2 t^2 \over 2520}$  & \tend
height20pt  & $A_4 (1^+, 2^I, 3^+, 4^I)$  &
$ \ i \, { ( \sqrt{2} g)^4 \over (4 \pi)^5 \ \eps} \ { \spa2.4^4 \over \spa1.2 \spa2.3 \spa3.4\spa4.1 } \ {s t \over 2} $   &
$ - { 3 s^2 + 2 s t + 3 t^2 \over 181440}  $  &
$ 0 $   &
$ -{2 s^2  + s t + 2 t^2 \over 2520} $  &
$ {2 s^2  + s t + 2 t^2 \over  2520} $ & \tend
height20pt  & $A_4 (1^-, 2^+, 3^I, 4^I)$  &
$-i \, { ( \sqrt{2} g)^4 \over (4 \pi)^5 \ \eps} \ { \spb1.2^2 \spa1.3^2 \over \spb1.2 \spa2.3 \spa3.4\spb4.1 } \ {t^2 \over 2 s} $   &
$ - { s ( 2 s + 3 t) \over 181440} $  &
$ { s u \over 240} $   &
$ -{ s (16 s + 13 t) \over 5040 }$  &
$- { s (5 s + 8 t) \over  5040  }$ &\tend
height20pt  & $A_4 (1^-, 2^I, 3^+, 4^I)$  &
$-i \, { ( \sqrt{2} g)^4 \over (4 \pi)^5 \ \eps} \ { 1 \over \spb1.2 \spa2.3 \spa3.4\spb4.1 } \ { s^2 t^2  \over 2 u} $   &
$ { u^2 \over 45360} $  &
$ { u^2 \over 240} $  &
$ { u^2 \over 420}$  &
${ u^2 \over 560} $  & \tend
}
}
}
}
\nobreak
\caption[]{The $1/ \epsilon$ infinities for the four-point one-loop gluon amplitudes in $D=10$.
\label{AmplitudesD10}
\small
\smallskip}
\end{table}

%
\begin{table}[t]
\hbox{
\def\tend{\cr \noalign{ \hrule}}
\def\t#1{\tilde{#1}}
\def\tw{\theta_W}
\vbox{\offinterlineskip
{
\hrule
\halign{
       &  \vrule#
        &\strut\hfil #\hfil\vrule
        &\strut\hfil #\hfil\vrule
        &\strut\hfil #\hfil\vrule
        &\strut\hfil #\hfil\vrule
        &\strut\hfil #\hfil\vrule
        &\strut\hfil #\hfil\vrule
       \cr
height15pt  &{\bf Amplitude}  & {\bf O F }   &
$\bf A_4^{[S]}$  & $\bf A_4^{N=1}$  & $\bf A_4^{[V-S]}$ & $\bf A_4^{[F-S]}$ &\tend
height20pt  & $A_4 (1^I, 2^I, 3^J, 4^J)$ &
$\ i \, { ( \sqrt{2} g)^4 \over (4 \pi)^5 \ \eps} $   &
$ { 33 s^3 + 7 s^2 t + s t^2 + 3 t^3 \over 362880}  $ &
$ { t u^2 \over 480} $ &
$ { 6 t^3 + 15 t^2 u + 26 t u^2 + 4 u^3 \over 10080} $ &
$ - { 6 t^3 + 15 t^2 u + 5 t u^2 + 4 u^3 \over 10080} $ & \tend
height20pt  & $A_4 (1^I, 2^J, 3^I, 4^J)$ &
$\ i \, { ( \sqrt{2} g)^4 \over (4 \pi)^5 \ \eps} $   &
$ { 3 s^3 + s^2 t + s t^2 + 3 t^3 \over 362880} $ &
$ { s t u \over 480} $ &
$ { 2 s^3 - 3 s^2 t - 3 s t^2 + 2 t^3 \over 5040} $    &
$ - { 4 s^3 + 15 s^2 t + 15 s t^2 + 4 t^3 \over 10080} $ & \tend
height20pt  & $A_4 (1^I, 2^I, 3^I, 4^I)$ &
$\ i \, { ( \sqrt{2} g)^4 \over (4 \pi)^5 \ \eps} $  &
$ { 13 s^3 + 3 s^2 t + 3 s t^2 + 13 t^3 \over 120960} $  &
$ { s^3 + 2 s^2 t + 2 s t^2 + t^3 \over 480} $  &
$ { 13 s^3 + 33 s^2 t + 33 s t^2 + 13 t^3 \over 10080} $  &
$ { 8 s^3 + 9 s^2 t + 9 s t^2 + 8 t^3 \over 10080} $  & \tend
}
}
}
}
\nobreak
\caption[]{The amplitudes in $D=10$ involving purely $(D-4)$
dimensional helicities.
\label{AmplitudesD10_con}
\small
\smallskip}
\end{table}

$$
\eqalign{
H^{[S]} = & - {( \sqrt{2} g)^4 \over (4 \pi)^5 \epsilon} \ {N_c \over 181440} 
\ \biggl( 4 H_1 + 2 H_2 + 6 H_3 + 15 H_4 - 12 H_7 \biggr)
\cr
H^{[F]} = & - {( \sqrt{2} g)^4 \over (4 \pi)^5 \epsilon} \ {N_c \over 90720} 
\ \biggl( 64 H_1 + 212 H_2 + 96 H_3 - 21 H_4 - 63 H_5 - 45 H_6 + 24 H_7 \biggr)
\cr
H^{[V]} = & - {( \sqrt{2} g)^4 \over (4 \pi)^5 \epsilon} \ {N_c \over 181440} 
\ \biggl( 628 H_1 + 332 H_2 + 564 H_3 - 147 H_4 - 63 H_5 - 99 H_6 - 48 H_7 \biggr)
\cr
H^{N=1} = & - {( \sqrt{2} g)^4 \over (4 \pi)^5 \epsilon} \ {N_c \over 960} 
\ \biggl( 4 \left( H_1 + H_2 + H_3 \right) - \left( H_4 + H_5 + H_6 \right) \biggr)
\cr}
\equn
$$

\noindent

The supersymmetric amplitude has, again, been calculated
previously~\cite{GSB}.  If we had been using a cut-off regularisation
scheme we would have expected a counterterm of the form $\Lambda^2
F^4$ and then compatibility with maximal supersymmetry would force the
$F^4$ to be of the unique form $t_8\cdot F^4$. In the case of
dimensional regularisation the counterterm is of the form $D^2F^4$ and
can be rewritten as
$$
 H^{N=1} =  - {( \sqrt{2} g)^4 \over (4 \pi)^5 \epsilon} \ {N_c \over 1920} \
t_8\cdot D^2F^4
\equn
$$
where
$$
\eqalign{ t_8 \cdot D^2 F^4 =& \ t_8^{a_1 b_1 a_2 b_2 a_3
b_3 a_4 b_4} \ D_{c} F_{a_1 b_1} D^{c} F_{a_2
b_2} F_{a_3 b_3} F_{a_4 b_4} \cr =& \ - 2 \left[ \Tr \left(
D_{e} F_{ab} D^{e} F^{ab} F_{cd} F^{cd} \right) + \Tr \left( D_{e}
F_{ab} D^{e} F_{cd} F^{ab} F^{cd} \right) + \Tr \left( D_{e} F_{ab}
D^{e} F_{cd} F^{cd} F^{ab} \right) \right] \cr & \ + 8 \left[ \Tr
\left( D_{e} F_a^{\ b} D^{e} F_d^{\ a} F_b^{\ c} F_c^{\ d} \right) +
\Tr \left( D_{e} F_a^{\ b} D^{e} F_b^{\ c} F_c^{\ d} F_d^{\ a} \right)
+ \Tr \left( D_{e} F_a^{\ b} D^{e} F_c^{\ d} F_b^{\ c} F_d^{\ a}
\right) \right] \cr =& \ \ 2 \ \biggl( 4 \left( H_1 + H_2 + H_3
\right) - \left( H_4 + H_5 + H_6 \right) \biggr) \cr}
\equn
$$
This is, necessarily, compatible with supersymmetry.  As one can see the
covariant derivatives, $D^c$, are contracted with each other and the tensor linking the four
field strengths, $F_{a b}$, is $t_8$.
The infinity for maximal supersymmetry can be written as 
$$ 
\sim st \Atree \times (s+t)
\equn 
$$
and extending this expression to the case where the tree amplitudes
are for external gluons and/or gluinos will generate the full quartic
component of the counterterm.  Again these counterterms are
non-vanishing and so for both the $D=8$ and $D=10$ cases a larger
symmetry group than supersymmetry is required to produce a finite
one-loop theory.

\vskip 0.3 truecm

\section{Conclusions}

In this letter we have evaluated physical on-shell amplitudes and
determined the counterterm structure for higher dimensional
Yang-Mills. The availability of specialised modern techniques for
calculating on-shell loop amplitudes makes this a more efficient
technique than the tradition approach of evaluating smaller point
off-shell functions.

Since higher dimensional Yang-Mills is a gauge theory with a
dimensionfull coupling constant we expect many of the structures we
obtained to arise in a similar manner for gravity calculations.  With
increasing dimension the complexity of the counterterm structure
increases significantly and the possibilities of obtaining a finite
predictive theory are more restrictive, probably requiring 
the existence of a much
larger symmetry group, such as the symmetries of string theory. 
Supersymmetry, in itself, is not enough to provide these symmetries.

\vfill\eject

\vfill\eject

\end{document}